\documentclass{article}

     \usepackage[preprint]{neurips_2019}

\usepackage[utf8]{inputenc} 
\usepackage[T1]{fontenc}    
\usepackage{hyperref}       
\usepackage{url}            
\usepackage{booktabs}       
\usepackage{amsfonts}       
\usepackage{nicefrac}       
\usepackage{microtype}      

\title{Recursion, Probability, Convolution and Classification for Computations}

\author{%
  Mircea~Namolaru
  \\
  CEA, Nanno-Inov\\
  \texttt{mircea.namolaru@cea.fr} \\
  \And
  Thierry~Goubier\\
  CEA, Nano-Inov \\
  \texttt{thierry.goubier@cea.fr} \\
}

\begin{document}

\maketitle

\begin{abstract}
The main motivation of this work was practical, to offer
computationally and theoretical scalable ways to structuring large
classes of computation. It started from attempts to optimize R code
for machine learning/artificial intelligence algorithms for huge data 
sets, that due to their size, should be handled into an incremental (online) 
fashion.   

Our target are large classes of relational (attribute based), mathematical (index based) 
or graph computations. We wanted to use powerful computation representations that 
emerged in AI (artificial intelligence)/ML (machine learning) as BN (Bayesian networks) and 
CNN (convolution neural networks). For the classes of computation addressed by us, and 
for our HPC (high performance computing) needs, the current solutions for translating 
computations into such representation need to be extended. 

Our results show that the classes of computation targeted by us, could be tree-structured, 
and a probability distribution (defining a DBN, i.e. Dynamic Bayesian Network) associated 
with it. More ever, this DBN may be viewed as a recursive CNN (Convolution Neural Network).
Within this tree-like structure, classification in classes with size bounded (by a parameterizable 
may be performed.   

These results are at the core of very powerful, yet highly practically algorithms for 
restructuring and parallelizing the computations. The mathematical background
required for an in depth presentation and exposing the full generality of our approach)
is the subject of a subsequent paper. In this paper, we work in an limited (but important) 
framework that could be understood with rudiments of linear algebra and graph theory. 
The focus is in applicability, most of this paper discuss the   
usefulness of our approach for solving hard compilation problems related to automatic 
parallelism.  
\end{abstract}

\section{Executive summary}
The underlying assumption of this work (and of automatic computation structuring) 
is that the programmer is not required to parallelize and/or manage the flow of data.
The industrial setting in which we worked forced us from the very beginning
to work with huge sets of multi-dimensional data, structured in arrays and 
graphs. Approaches as polyhedra model, that computationally and 
mathematically suffers from serious scalability problems (i.e. works only 
for a small number of dimensions) clearly were not appropriate. 
We needed to embark ourselves in a risky and difficult path, searching 
for suitable solutions to the challenging problems of this project  
and of automatic parallelism in general. 

We think that our work is a big step forward for automatic computation structuring.
It was enabled by our dataflow and probabilistic view of computations. 
 This is based on successful generalization and synthesis of powerful computational models 
emerging from AI (as BN, Bayesian Networks) and ML (as CNN, Convolution Neural Network).

In a sense, it is the confirmation of  
the main author belief that the key 
to automatic programming are metric characterizations of graphs and relational 
structures.      
The work, described in \citet{Nm}, generating metrics via aggregations 
is continued and generalized by this work.
   
Our approach target large classes of computations, and is theoretically and  
computational scalable.
The computations are represented as graphs, to which a probability distribution,
or other metrics, may be associated. We note that in our approach, linear 
computational models (affine partitioning framework, or polyhedra model) emerging 
from compilation parallelization appear as 
special cases, using subsets of the transformations considered by our 
approach. 

This technical report describes our approach in a simplified 
setting. Its aim is to present a non-demanding and rather intuitive technical 
presentation, supporting the (admittedly far reaching) claims regarding the 
potential of our computation structuring work. The key to our success is a  

Inherently, due to the scope of this work, we need to limit ourselves to a 
reduced mathematical framework and simple examples from the most significant 
areas benefiting from our data-flow driven computation structuring: ML (convolution 
and neural networks), AI (Bayesian networks, relational and database computing), 
HPC, graph computations, software engineering (programming patterns), legacy code 
and refactoring.

Future work will include an in-depth mathematical presentation of  
our work showing its full generality and applicability, and an accompanying powerful 
automatic structuring tool (that we hope to set new standards in the industry).    
\section{Technical introduction}
The main ideas of our approach are presented below:

We address the problem of decomposing Cartesian product spaces (further denoted 
as index spaces), central to relational (attribute based), mathematical 
(index based) and graph computations. This boils down to enumerations of points
of these spaces, defining subsets with desirable properties as ``linear independence'' 
(``parallelism'') or ``recursion''. Some computation constraints
(denoted as dependencies) should be preserved, ensuring that the computation
result remains the same. 

Index spaces are viewed as oriented acyclic graph (dags) having a root (origin).
The enumeration of its vertexes is done by considering paths from origin to
vertexes. A time index is added to the initial indexes. The enumeration assigns 
to each vertex, a time value defined by the length of the path used to reach the vertex. 
Only path lengths (or equivalently time values) that are sums of powers of two
($2$-adic numbers) are used to 
enumerate the graph vertexes, defining an enumeration defined by a tree 
recursive order. 

The computations are defined by sets of formulas over indexes spaces,
defining dependencies (i.e partial order relations). The enumeration algorithms  
present by us, preserve these order relations. For some computations, temporary space 
need to be used to comply with this requirement. The management of this space is
done by the enumeration algorithm, and it is an integral part of it).  

Underlying our methods, there is a geometrical view, computationally expressed 
via integrals (summations) over measures (\citet{Co}) defined by finite (discrete) groups
(\citet{Sh1}). 
In this mathematical framework, strong connections between DBN, 
recursive CNN and classification emerge. This is rather expected as 
probabilities (normalized measures), convolution products (and Fourier transformations)
and partitions (measurable sets) are all "clients" of the measure theory. 

In this paper, we restricted ourselves to a limited mathematical framework, requiring 
only rudiments of linear algebra and graph theory. We use an informal style of 
presentation, there are no proofs, and no attempts to present the problems and solutions 
in a general mathematical setting. 
 
Our work will be discussed in its full generality in a 
subsequent paper. Here we focus on the applicative aspects, relevant to computation 
restructuring (and parallelization), trying to convey the main ideas and potential of 
our approach. 
\subsection{Overview}
Our targets for restructuring are relational (attribute based), mathematical 
(index based) and graph computations. We present the detailed plan of the paper, 
for each section.
\begin{itemize}
\item Index spaces. The target computation, expressed as an map-reduce computation, 
defined by a set of equalities/formulas over index spaces.
Legal formulas for the indexes and the dependencies defined by them.   
\item Index space and the lattice (graph) defined by it. Lattice points enumeration 
via paths. The lattice coloring and partitions induced by coloring. 
\item The time-index space, defined by addition of a time index to index spaces.
"Clock", the time "unit" and its graduations. 
Clocks defined by sequences of powers of two.  
\item Enumeration of time-index space using the clock structure and mapping of 
indexes to clocks. Enumeration of clock products.    
\item The cube, as an "unit" for the index space. The relations between "clocks" and "cubes".
Time-index space as a recursive tree (with a sub-tree as the 
recurrent pattern). The accumulator pattern.  
\item Complex dependencies, stencils, transpose, matrix multiplication. Enumeration 
and allocation of temporary spaces for them.
\item Sparse index space and graphs enumeration. Dynamic vs. static scheduling.
Depth-first vs. breath first enumeration.       
\end{itemize}

\subsection{Contribution. Compilation}
Inferring recurrent computational structures from iterative ones is a difficult
problem, relevant for artificial intelligence, machine learning, database or
automatic code restructuring (parallelization). 

Extension of powerful computation representations from AI and ML, enable representation  
of indexes space as recursive tree. This ensure a tree-width complexity of index spaces 
enumeration. 
In our approach, the parallelism is based on a convolution paradigm, replacing the
influential Lamport parallel hyperplane paradigm.
 
The representation of computation as map-reduce pattern, unifies linear (loop based),
stencil and graph computation. We provide new ways of defining dependencies, and efficient ways
of preserving them by using temporary memory space for partial results. In our approach the 
management of temporary space is an integral part of index space enumeration. As result,
the parallelization depends on the temporary space size.  
  
The structuring (parallelization) uses a graph and a path view of index spaces. Our
approach is suitable for both sparse and index representation of index spaces. Graph coloring 
is used to define partitions of the index space, the coloring being a model for hardware 
resource allocation. It defines fixed buffers in which the index space points are brought.  

In our approach, the enumeration is view as integration (summation). 
The transformation performed are similar to the one performed for Lebesgue 
integral (Fubini, convolution product, etc) (\citet{St}). The initial computations 
is specified as a multiple integral over product spaces.

The separation (induced by coloring) make possible the transformation of this integral 
into sequences of nested integrals, where the function
integrated is a recurrent computational pattern.

To help the understanding of examples, an equivalent loop based notation is used 
instead of the integral one.  

\subsection{Contribution. AI and ML}
The use of powerful computation representations as BN and CNN, that emerged in AI 
and ML and are behind impressive results, is in our opinion a major 
contribution of this work. It make possible to use (or adapt) the rich body of existent 
work already done. 

But as mentioned, we needed to expand existent results for these representation. 
Our contributions are enumerated below, and are not discussed further in the paper 
(albeit our examples with a different terminology would illustrate them).  

We are able to  infer DBN on index spaces. 
BN (\citet{Da}) are extended to work with indexes with an arbitrary 
number of values (and not only with boolean indexes). Index spaces are structured 
as recurrent BN over time (DBN). 

This DBN is also a recursive CNN (\citet{Go}) whose convolutions
represent conditional independence. The distinction between a
neural network (computing a function) and a Bayesian network (computing a
probability distribution) is blurred. Moreover, a classification of
index (product) spaces, in classes with size bounded by a parameterizable threshold
is defined by the coloring. 

\section{Computation. The product view}

\subsection{Indexes and index spaces}

We denote by small\footnote{The use of small letters for variables and upper letters for
indexes comes from logic programming (this is the Prolog convention,
so our formulas may be easily transposed to Prolog programs
or vice-versa)} letters $x,y,z \dots$ variables.

A variable denotes a set of values that could be enumerated
(or indexed). An index $I$ denotes the set $\{1,2,\dots\}$. We may view
variables as a function $x:I = \{1,2,\dots\} \rightarrow
  Vals = \{x(1),x(2),\dots\}$ where $x(K)$ denotes the $K$ value of the
  variable $X$.

We may consider multi-dimensional variables, defined on index spaces (i.e.
Cartesian products of indexes). For instance $x:I \times J \times K \rightarrow
x(I,J,K)$. Enumerating the values of such multi-dimensional variables are
crucial for computation structuring, and complex enumeration exhibiting some
desirable properties could be viewed as the main subject of our
paper.

\subsection{Formulas. Map-reduce patterns}
The computations are specified by a number of formulas (functions) specifying
computational patterns for variables, that should be applied on all values of 
the index space upon which variable are defined.   

For instance $a(I,J)+= b(I,K)*C(K,J)$ for all the the values
(tuples) $(I_{i},J_{j},K_{k})$ of the index space $I \times J \times K$. This
is example is matrix multiplication.
Another example is $f(M,N) = g(M,N)*h(P,Q)$ for all the tuples 
of the index space $M \times N \times P \times Q$.
And another one (defining a stencil computation on a grid) is
$grid(I,J) = grid(I+1,J) + grid(I,J+1) + grid(I+1,J+1)$, for all tuples
of $I \times J$.

The underlying indexes in a formulas, define polynomial formulas, 
that should be applied for all tuples of the formula index space. For the 
examples from the preceding subsection, we obtain
$IJ = IJ + IK^{2}J$, 
$MN = MNPQ$ and $IJ = (I+1)J + I(J+1) + (I+1)(J+1)$. 

This is similar to the ``reduce-map'' pattern, specifying a computational
pattern ("map" part) that should be performed on all the points of a domain. The 
actual way of enumerating the domain points ("reduce" part) is left unspecified.

There is a major differences, for us the computational patterns are   
formulas $Res = Operands$ between indexes, where $Res$ is the set 
of indexes appearing in formula result, and $Operands$ is the set of indexes appearing in
formula operands. The pattern is a reduction from a $n$-space (defined by operand indexes) 
to a $m$-space (defined by result indexes) with $m \le n$. 

We assume that the underlying memory spaces indexed by $Res$ indexes and by $Operands$ indexes 
are the same. The same memory location is used by $Res$ indexes (for storing a result,
a "def") and also by $Operands$ indexes (for loading a result, an "use"). 
The formulas specify given associations between ``defs'' and ``uses'', denoted as
dependencies. The enumeration should comply with this association in order to preserve
the computation results. A "def" should precede its "uses", and should stay alive (i.e. 
its memory location should not be overridden), till all its uses have been performed. 

In this "map-reduce" based approach, stencil, loop based or graph based computations 
are unified. This is done by associating a sequential enumeration of the index space 
to the map-reduce pattern, compliant with the dependencies. The computation restructuring 
may be view as replacing the sequential enumeration with another one with desirable 
properties ("parallelism", "recursion"), while complying with the dependencies.
 
We note that many mathematical (loop based) computations assume a lexicographic order
of indexes, and are already expressed via a sequential enumeration. Like-wise, graph 
computations using a depth-first order of vertexes. In general for any index space 
and any dependencies defined by formulas, a sequential enumeration 
(compliant with dependencies) may be inferred. As further discussed additional 
memory space may be required for partial results.
\subsection{Legal formulas, Dependencies}
We will impose the following syntactic restrictions on the formulas 
defining the computational pattern.
The formulas considered define functions whose domain are the result index
space, and the range are the operands index space.   

The $Res$ part is formed by an index product (i.e. monomial), each index 
with exponent one. The indexes may be equal (so $II$ is legal, but 
$I^{2}$ is not). The $Operands$ part is formed by a sum of factor products,
where a factor is an index plus an integer (denoted as displacement) with 
exponent one. For instance, for stencil the formula is 
$IJ+=(I+1)J+(J+1)I+(I+1)(J+1)$. 

The order of factors (indexes) in the products, as well as the paranthesis,
is the one defined by the indexes relations/arrays in the computation 
pattern. For matrix transpose defined by $A(I,J) = A(J,I)$, we get the 
formula $IJ=JI$. Likewise matrix multiplication yields the formula
$IJ=IJ+(IK)(KJ)$  Several formulas may be used, where a $Res$ may appear 
as a factor in an  $Operand$ product, possibly adding displacements to the 
indexes in $Res$.  

The parenthesis structure, and the displacements define cyclic structures 
between indexes appearing in the formulas result and formulas operands,
defining an order (i.e. dependencies) between their values.
A sequential enumeration (i.e. the index space is view as a sequence) 
compliant with the dependencies is defined, possibly requiring additional 
temporary space, to preserve overwritten values (another "def" use their
memory location) that still have further uses.   

For this sequential enumeration, the temporary memory size required 
is a constant. Given a temporary memory size threshold (greater or equal with
this constant), alternative enumerations compliant with the original data 
dependencies, may be inferred. The parallelism achieved via alternative 
enumerations, is function of the temporary memory size.   
The basic ideas are exemplified for stencil computation, matrix transpose and
matrix multiplication.    

\subsection{Legacy code. Sequential enumeration. Lexicographic and depth-first order}
Many computations are not expressed via a map-reduce pattern, and specify an 
enumeration of the index space. 
The sequential enumeration corresponds to the lexicographic 
order for a loop based computation, and to a depth-first order for a graph based 
computation. In the first case, the sequential
enumeration is specified as an iterative computation, and in the second case as a 
recursive one. 

The sequential enumeration, requires the minimum of temporary space for complying 
with dependencies. It is also the enumeration with minimum parallelism (having no 
parallelism at all). Alternative enumerations increase the parallelism, but also 
increase the temporary memory size required for complying with the dependencies.
An enumeration cannot maximize the level of parallelism and minimize the temporary 
memory, but only to set a "good" balance (in terms of hardware resources) between 
them.  
 
A similar relation exists between scheduling (exploiting instruction level 
parallelism) and register allocation, triggering an approach 
where scheduling and register allocation are viewed as a single 
transformation.
Our approach may be view as a generalization of such algorithms.
   
Many mathematical computations are formulated using a sequential enumeration, 
and sequential programs are a straightforward encoding of such formulations.
For these programs, the memory space indexed by the result indexes is different 
from the memory space indexed by operand indexes. In this way, complex alias 
analysis, with results usually too conservative to be useful, is avoided. 
Legacy code (or usual mathematical encodings) are presented in this way.

There are computations, where the result and computation indexes are indexing the
same memory space (stencil and grid computation, matrix transpose etc). These 
computations are defined via a map-reduce pattern, and a simple (but inefficient
in terms of temporary memory required) a sequential enumeration possible unfolding 
a formula via several ones, is first inferred.
From the sequential enumeration, alternative "parallel" enumerations are inferred from the sequential one. 
a scheduling with maximum parallelism, for a given size of temporary memory
(defined by the underlying memory hierarchies).  

The sequential enumeration  enumerate the lattice points using only forward 
and no convolutions. In compilation terms, the enumeration has maximal temporal 
locality, and no parallelism. The convolutions, use backward edges, and the convolution levels introduce (bounded) subsets of parallel points. The temporal locality is decreased and the parallelism increased. The maximum parallelism is achieved when all the computation indexes
are parallelized, i.e. reached via convolutions.

\subsection{Coloring}
It is one of the tenets of our work that deciding what are the intermediate results,
is an integral part of the enumeration. This may be view as a form of computation 
partitioning similar with tiling or blocking, but the intermediate results   
may be related to the hardware resources available, modeled as colors. There is an 
underlying coloring of the index space with a finite number of colors associated to 
an enumeration.    

Intuitively, the idea of recursion and classification (in a finite number of 
classes) are equivalent, as one implies the other. Classification and measure are 
also equivalent, as we may view the classes as covering (``measuring'') the space. 
This intuitive notions, may be formally expressed using the theory of groups.
But as a difference from traditional classification, in our approach the classes 
(i.e. the colors) are groups, endowed with a structure defined by a composition operation. 

\section{Computation. The lattice (path) view}
We will transforming the product space, into a lattice (oriented graph).
The enumeration problem is translated into a graph paths problem. 
This made possible, as exemplified in a further section, to use the same 
algorithms for both the linear (index based) and graph computation. 
   
\subsection{Lattice}
We consider a product space defined by the indexes $I_{k}$.
An index $I_{k}$ values are of the form $\{1,2,\dots,N_{k}\}$, where 
$N_{k}$ specifies the number of index values.
We associate to each index $I_{k}$ a vector $e_{k}$ of a standard 
orthonormal base. An index $I_{k}$ is view as a vector space, having 
as base the vector $e_{k}$, i.e. $I_{k}=<e_{k}>$.
 
The direct sum of two vector spaces $L,K$ is defined as 
$L+K=\{l+k|l \in L,k \in K\}$. The index space is the direct 
sum of $<e_{1}>+<e_{2}> \dots$. It is the lattice generated by the 
vectors $e_{i}$, whose elements are linear combinations 
$a_{1}e_{2}+a_{2}e_{2}\dots $, with $a_{k} \in I_{k}$.

By introducing an origin (transforming the lattice from
a vector space to an affine one), a lattice point is associated with 
oriented paths formed by the vectors $e_{i}$ connecting the origin
with the point. A path is represented by the lattice vertexes, i.e.
the start vertex of the path and the targets of the vectors $e_{i}$
composing it. The length of the path is the number of its vertexes. 

During the enumeration process we want a lattice point to be 
enumerated only once, so a single path (from the set of paths 
reaching it from the origin) should be considered.
In this view, is easy to see why the lexicographic (or depth-first)
order is inefficient. Some of the paths are traversed over and over
again in order to reach all the lattice points.  

\subsection{Time and clock}
Important for our approach is the introduction of a new 
time index for the lattice points. Every lattice point receives
an additional time index. The $n$-dimensional index
product space is transformed into a $(n+1)$-dimensional time-index space.
This may be viewed as the embedding of the lattice in a projective 
space (\citet{Sh}).   

The time dimension is structured around time units, representing disjunct
segment of size $2^{k}$. The powers of two $2^{l}, l<k$, define the graduations
of a time unit, i.e. its structure that is denoted as a ``clock''.
Alternative structuring of the time index and its associated
``clock'' are possible, but not discussed or used in this paper. Here, the
clock graduations (structure) is synonym with a sequence of values consisting of 
powers of two. Each value of the sequences 
is obtained by multiplying the precedent one by a power of two, the step of the sequence. 

The product space enumeration is driven by the time and its associated
"clock" structure. It may be view as a scheduling of lattice points, associating
time index and its clock graduations to lattice  points.

\subsection{Paths}
In the  lattice view, the lattice points are enumerated by using paths 
constructed from the lattice edges. The decomposition of the lattice is 
driven by length of paths from the origin to lattice points. In the lexicographic order 
there are only acyclic paths, moving forward, i.e. defined by linear combinations 
of lattice points with non-negative coefficients. The convolutions introduce 
parallelism (and paths using backward edges) still preserving 
the lexicographic order.  

We will consider paths whose lengths are associated to a given "clock" (i.e. 
are sequences of powers of two, $2^{l}$ with $l \le k$, where $k$ 
is a given parameter). Every path in the lattice is structured as a composition 
of such paths. Any lattice points is associated to a linear combination of 
powers of two, where the coefficient of $2^{k}$ is a non-negative integer, 
and the coefficients of $2^{l},0 \le l<k$ are zero or one.  

This association of the lattice points with linear combinations 
of powers of two defines a coloring. To every exponent $l$ of a power $2^{l}$ is 
associated a different color. The color of the smallest exponent appearing
in the linear combinations associated with a point is assigned to the point. 

By associating lattice points to time index and its graduations, 
we may view enumeration as transitions in a time-index space, and associate
 a probabilistic distribution to the enumeration.  

\section{Computation. The enumeration view}
The sequential order is parallelized by structuring the lattice as a tree
structure (a graded) order in the time-index space. The parallelism is 
provided by convolutions induced by the time index and its corresponding 
clock.  
 
\subsection{Time and clock skeleton}
First, the enumeration of a clock points, i.e. the points spawned by "clock" 
(graduations) is considered.   
This provide the basic underlying computational 
pattern for our approach. The time and its graduations define a skeleton 
for enumeration. The extension to lattice points is done via a mapping (discussed 
in the next subsection) of indexes to "clock" graduations (i.e. powers of two)

We consider a $3$-clock defining an unit time of $8$ points with the
graduations $T=2^{2}=4,TX=2^{1}=2,TY=2^{0}=1$. Enumerate the points
spawned by the clock graduations (i.e. linear combinations of 
$T,TX,TY$ with coefficients in $\{0,1\}$.

\begin{verbatim}
   for (T=0;T<8;T+=4)
     for (TX=0;TX<4;TX+=2)  
       for (TY=0;TY<2;TY+=1)
         (T,TX,TY)
\end{verbatim}
 
The points $(T,TX,TY)$ are 
enumerated in the order $(0,0,0),(0,0,TY),(0,TX,0),(0,TX,TY),$
$(T,0,0),(T,0,TY),(T,TX,0),
(T,TX,TY)$  that is  the lexicographic (or depth-first) order. 

The $(T,TX,TY)$ indexes define a tree, partitioning the time
index in subsets with size powers of two. The lexicographic order
enumeration linearize this tree, and don't take advantage of the 
existing parallelism in a tree structure.

\subsection{Convolutions}
We consider two points $a,b$ that are reached from the
origin by vectors (paths) of same length $pa,pb$. We want to use $pa$ 
to reach $b$ using the existent vectors (paths) system, and eliminate $pb$. We 
do this by considering the composition of paths $pa+pb$, and reversing the 
direction of $pa$ to reach $b$. The point $b$ is reached via a convolution of 
$a$ defined by the path $(pa+pb)-pa$. We may have different levels of convolution, 
for instance in the previous example $pa+pb$ may also be eliminated and reached via 
convolutions.

The convolutions use backward edges, and don't preserve the lexicographic order 
requiring to advance only in forward paths. The convolutions structure
the space of linear independent points, by replacing the lexicographic order
by a graded order compatible with it. The graduations of this order are the
convolution levels (consisting of points with same length from origin, hence linear 
independent). The convolutions provide a way of performing the coloring incrementally 
during the enumeration, by coloring the points at a given convolution level with the 
same color.

For every $k$ clock its points may be enumerated via $k$-level convolutions.
The introduction of convolutions is done via a set of linear transformations, 
passing from paths with length powers of two, to paths with length sum of 
powers of two.

\begin{verbatim} 
  T->id
  TXN->TX+T
  TYN->TXN+TY->T+TX+TY
\end{verbatim}

The points $(T,TXN,TYN)$ represent paths whose length are  
sums of powers of two. Other paths of same length used
for enumerating may be eliminated, and convolutions defined 
by the points $(T,TX,TY)$ are used to reach the same points 
as before.
 
The enumeration of the $3$-clock from the previous examples via 
 $2$-level convolution is shown below.   
 
\begin{verbatim}
   for (T=0;T<8;T+=4)
     for (TXN=T;TXN<T+4;TXN+=2)  
       for (TYN=TXN;TYN<TXN+2;TYN+=1)
         (T,TXN-T,TYN-TXN) 
\end{verbatim}

In this form the loop structure may be serialized
using a sequence of nested $enum$ functions, similar 
to an integral. The bounds of this function are set using  
the index and step of the previous function in the sequence.
   
\begin{verbatim}
   enum(T,4,[0,8))
     enum(TXN,2,[T,T+4))
        enum(TYN,1,[TXN,TXN+2))
           (T,TXN-T,TYN-TXN) 
\end{verbatim}  

Similar constructions may be done for any $k$-dimensional clock,
with any advancement rate. 
\begin{verbatim}
   enum(INDEX,STEP,BOUNDS)
     enum(INDEX1,STEP1,BOUNDS1(INDEX,STEP))
        ...
           formula
\end{verbatim}

The time index is structured as an enumeration of time "units" 
represented by a clock. The enumeration loop is itself a
a clock, whose rate step is defined by the size of the
"unit" clock. In this way levels of nested clocks may be defined,
adding additional graduations on the time index. 

\begin{verbatim}
  We assume that the size of the time dimension is 64.
  
  TG->id
  TGXN->TGX+TG
  TGYN->TGXN+TY->T+TX+TY
  
  /* enumeration clock of the clock units
     enumerating its origins. */
  for (TG=0;TG<64;TG+=32)
     for (TGXN=TG;TGXN<TG+32;TGXN+=16)  
       for (TGYN=TGXN;TGYN<TGXN+16;TGYN+=8)
\end{verbatim}

\begin{verbatim}
  /* Enumeration of the clock unit. Its
     origin (index) T should receive 
     all the points of the outer clock. */
     
     The point T (the origin) receive as values the
     points of the outer clock. 
     
     T->TGYN
     TXN->TX+T
     TYN->TXN+TY->T+TX+TY
     
     for (T=TGYN;T<TGYN+TGXN+TG+8;T+=4)
       for (TXN=T;TXN<T+4;TXN+=2)  
         for (TYN=TXN;TYN<TXN+2;TYN+=1)
           (T,TXN-T,TYN-TXN) 
\end{verbatim}

The time index is represented as the product of the ``unit'' clock
with the outer clock. The points of each of these clock may be 
enumerated via convolutions. There are different alternatives for decomposing 
the index space in a product of smaller clocks (i.e. factorization).
For our example, the time space may be considered as the product of six $1$-clocks,
two $3$-clocks, or three $2$-clocks. Examples are provided further in the paper.   

\subsection{Mapping of lattice points}
The lattice indexes are mapped to graduations of $k$-clock, i.e. powers of two,
defining the time-index space. In this space the step of an index is the power of 
two to which is mapped. In the time skeleton (template), the 
clock graduation indexes ($T,TX,\dots$ in previous examples) are replaced by the 
lattice indexes to which were mapped.

In the lexicographic order the paths were composed from segments with the same
length (two points). The mapping enables the use of segments with length different 
powers of two, having good compositional properties. 

The next example is provided by the formula $f(M,N)=g(M,N)h(P,Q)$.
In the lexicographic order the computation is expressed as:

\begin{verbatim}
  for (M=0;M<MS;M++)
    for (P=0;P<PS;P++)
      for (N=0;N<NS;N++)
        for (Q=0;Q<QS;Q++)
          f(M,N)=g(M,N)h(P,Q)
\end{verbatim}

We may associate the indexes $M,N,P,Q$ to powers of two defining a $4$-clock 
via the mapping $M \rightarrow 2^{3}=8$,$N \rightarrow 2^{2}=4$,$P \rightarrow 2^{1}=2$,
$Q \rightarrow 2^{0}=1$. Each index is mapped to a different clock graduation
(power of two).

We assume that the sizes $MS,NS,PS,QS$ of the indexes $M,N,P,Q$ are all $2$. After 
the mapping to  the corresponding clock skeleton is performed, we get the loop 
structure:

\begin{verbatim}
   for (M=0;M<16,M+=8)
     for (P=M,P<M+8;P+=4)
       for (N=P;N<P+4;N+=2)
         for (Q=N;Q<N+2;Q+=1)
           f(M,N-P)=g(M,N-P)h(P-M,Q-N)
\end{verbatim}

The indexes in the formula in the body of the loop should be 
reverted from their values in the time-index space, to their corresponding value 
in the index space. This is done by "undoing" the effect of mapping, and dividing 
an index by the power of two to which was mapped.

\begin{verbatim}
   f(M/8,(N-P)/2)=g(M/8,(N-P)/2)h((P-M)/2,Q-N)
\end{verbatim}

We are interested to increase the level of parallelism, by increasing the 
points at a given convolution level. This is done by mapping several 
indexes to the same clock graduations (power of two). We will consider 
a $2$-clock with step 4 defined by 
$(16,4)$, and the mapping $M,P \rightarrow 8$ and $P,Q \rightarrow 2$..
The resulting loop structure is represented as the product of two $2$-clocks,
each clock being enumerated via $1$-level convolutions. One of the indexes 
from the set of indexes mapped to the same clock graduation is chosen as a 
representative. The rest of the indexes from the set are reached via 
convolutions defined by 
this index.

\begin{verbatim}
for (M=0;M<16,M+=8)
  for (P=M,P<M+8;P+=4)
    for (N=P;N<P+4;N+=2)
      for (Q=N;Q<N+2;Q+=1)
        f(M,N)=g(M,N)h(P-M,Q-N)
\end{verbatim}

The indexes in the formula from the body of the loop should be divided by 
the powers of two to which were mapped. 
\begin{verbatim}
f(M/8,N/2)=g(M/8,N/2)h((P-M)/4,Q-N)
\end{verbatim}

We note that for this example we may increase the parallelism by adding an additional
$3$-clock structure $(TMP,M,N)=(32,8,2)$, reaching $M$ and $N$ via $2$-level convolutions,
and $P$ and $Q$ via $3$-level convolutions.   
\begin{verbatim}
for (TMP=0;TMP<32;TMP+=16)
  for (M=TMP;M<TMP+16,M+=8)
    for (P=M,P<M+8;P+=4)
      for (N=P+M;N<P+M+4;N+=2)
        for (Q=N;Q<N+2;Q+=1)
          // f(M-TMP,N-M)=g(M,N-M)h(P-M,Q-N)
          f((M-TMP)/8,(N-M)/2) = g((M-TMP)/8,(N-M)/2)h((P-M)/4,Q-N) 
\end{verbatim}

The indexes in formula from the body of the loop should be 
divided by the powers of two to which were mapped.
\begin{verbatim}
  f(M/8,(N-TMP)/2)=g(M/8,(N-TMP)/2)h((TMP-M)/4,TNQ-N-TMP)
\end{verbatim}
 
\section{A geometric interpretation}
In our approach, the computation partitioning is defined by the partial results
used by the computation. A formula (of a map-reduce pattern) may be 
unfolded into a set of formulas. Between these formulas there is a partial (acyclic) 
order, the result part of a formula being used in the operand part of other formulas. 

\subsection{Computation vectors and data flow}
The computation formula may be associated with a set of paths (vectors), 
the computation vectors, defining the directions of computations. Reaching the 
lattice points via paths formed by these vectors is in a sense optimal, 
as the computation follows exactly the flow of data. Practically, the use of 
computation vectors may be problematic. The computation vectors may not "fit" well 
together (i.e. cannot be expressed one in terms of another),
and lead to long vectors (paths) defining reuse distances too big for 
underlying architectures. 
  
In our approach, we work with a set of vectors, whose lengths are powers of 
two (bounded by a given threshold). These vectors have good compositional properties 
and provide an approximation of paths defined by computation vectors. These vectors 
are "close" to computation vectors, the deviations from the exact data flow 
being compensated by their properties. For index computations only multiplication 
and division with powers of two, that could be implemented efficiently via shifts 
on most architectures are used.  

The enumeration using these vectors (paths) should be compatible with the 
lexicographic order. The dependencies structure may require storage of temporary results
in order to ensure this requirement. The management of these temporary 
results is part of the enumeration algorithm, and is based on the coloring (that 
as previously discussed is related to convolutions and parallelism).   

\subsection{Cubes and clocks}
The lattice may be structured around units defined by "cubes". A $k$ cube is defined
as the Cartesian products of $k$ segments having the same size. The relations between 
clocks (time units) and cubes (lattice units) are discussed next.

The clock points spawned by linear combinations
with coefficients $\{0,1\}$ of a $k$-clock graduations (powers of two) may be 
associated with the points of a $k$-cube defined by $k$ segments $\{0,1\}$. 
This $k$-cube points are the $k$-tuples formed by $0$ or $1$. 

The $l$ vertex of a $k$ cube is the $k$-tuple with one in the
$l$ position, and zero in all the others. We map between between a $k$ cube
vertexes and a $k$ clock graduations, by associating to the $l$ vertex
the power $2^{l}$. More generally to a $k$-cube with sides of length $p$, 
corresponds a $k$-clock with rate $p$. The $k$-cube points are
the same as the $k$-clock points. 

We exemplify this for the clock defined in the time skeleton subsection
(that is reproduced here) and its corresponding cube. The loop structure for the 
cube is shown below. 
\begin{verbatim}
Cube of volume 8, defined by product of tree segments
[0,1]. Lexicographic order enumeration.     
for (T=0;T<2;T+=1)
  for (TX=0;TX<2;T+=1)  
    for (TY=0;TY<2;TY+=1)
      (T,TX,TY)
\end{verbatim}
    
We consider the transformation of indexes $(T,TX,TY) \rightarrow$ to $(4T,2TX,TY)$.
This is the mapping, setting the steps of $(T,TX,TY)$ to $(4,2,1)$.
The resulting loop structure enumerate the points of the corresponding clock.
The parallelism introduced by the mapping to the time index and its clock, may be view as 
time parallelism.
  
\begin{verbatim} 
for (T=0;T<8;T+=4)
  for (TX=0;TX<4;TX+=2)  
    for (TY=0;TY<2;TY+=1)
      (T/4,TX/2,TY)
\end{verbatim} 

The transformation below, introduce convolutions as already discussed in
the previous section. The time parallelism is enhanced with spatial
parallelism defined by convolutions.  
\begin{verbatim}
  T->id
  TXN->TX+T
  TYN->TXN+TY->T+TX+TY
\end{verbatim}

\begin{verbatim} 
for (T=0;T<8;T+=4)
  for (TXN=T;TXN<T+4;TXN+=2)  
    for (TYN=TXN;TYN<TXN+2;TYN+=1)
      (T/4,TXN-T/2,TYN-TXN)
\end{verbatim}

Geometrically, the mapping between a cube and a clock is the representation of
a cube by its main diagonals (a main diagonal connects the cube origin with
one of the cube vertexes). The clock representation defines a top-down
enumeration of a cube points by recursively expressing larger diagonals in terms of
smaller ones. This enumeration defines a recursive tree decomposition
of the cube points.
Viewing the cube points as the nodes of a tree is at the core of the lattice points
enumeration.

The $k$ cube points represents all the paths defined by the $k$ segments,
where each segment may have at most an occurrence). The $k$ clock associated
to it, use vectors with length powers of two (the main diagonals) to construct
paths reaching the lattice points. A $k$-clock with rate $p$,
and a $k$-cube with side $p$ are different representation of a set of 
$p^{k}$ lattice points, using different enumeration strategies.

\subsection{Product of cubes/clocks} 
We consider the product of two $k$-cubes with same side $p$.
It may be represented as a $2k$ cube with side $p$, as a $k$-cube with side 
$2p$. Depending on $k$ different factorization as a product of smaller
cubes are possible.   
Similar considerations holds also for the product of two $k$-clocks
with same rate $p$.

We exemplify by representing a 6-clock (32,16,8,4,2,1) with rate two,
as a 3-clock (32,8,2) with rate 4. The rate of this clock may be view 
as a 2-clock (2,1) with rate 2, and the 6-clock as the product of a 
3-clock with a 2-clock.
\begin{verbatim}
  for (TG=0;TG<64;TG+=32)
     for (TGXN=TG;TGXN<TG+32;TGXN+=16)  
       for (TGYN=TGXN+TG;TGYN<TGXN+TG+16;TGYN+=8)
         for (T=TGYN;T<TGYN+8;T+=4)
           for (TXN=T+TGYN;TXN<T+TGYN+4;TXN+=2)  
             for (TYN=TXN;TYN<TXN+2;TYN+=1)
               (TG/32,(TGXN-TG)/16,(TGYN-TG)/8,(T-TGYN)/4,(TXN-TGYN)/2,TYN-TXN). 
\end{verbatim}

\subsection{Accumulators, unfolding and parallelism}
Factorizations of a clock/cube, are expressed via clock graduations and convolution 
levels defined by them. 

The clock graduations represent the temporal locality, defining the points reached via 
forward vectors (paths) from the origin, while points reached via convolutions represent 
the parallelism. In this view, different factorization convert temporal locality in 
parallelism (and vice-versa). It is not possible to 
improve both at the same time, only to search for the most appropriate balance between them 
for a given computation and hardware configuration.

In the mapping done in the previous example, result indexes were mapped to
clock graduations, and the rest of indexes were reached via convolutions.
An alternative to this mapping, is to reach result indexes via 
convolutions of indexes not appearing in the result. These indexes 
(that may be operand indexes, or additional temporaries indexes) 
are used to accumulate partial results of an accumulator over the lattice points.
To any lattice enumeration may be associated an accumulator defining measures 
on the lattice points. Normalized this measures may define a probability 
distribution that is function of initial values associated to lattice points.

Additional formulas are added by "unfolding" (or unrolling) of indexes,
resulting in formulas using groups of consecutive index values. The unfolding
of an index increases its rate, as part of its values are now exposed in
formulas.     
In case when the index is an accumulator, the parallelism is increased as 
the computation is performed in parallel for unfolded values of the 
accumulator. 

We consider the case of a $3$-cube of size $2$, equivalent with the $3$-clock $(4,2,1)$.
The computation is required to sum all values indexing a $3$-dimensional
lattice, $S+=a(T,TX,TY)$. A new dimension $TMP$ with size $2$ is introduced, and associated
with the result $S$, that is transformed in a two-dimensional variable $s(TMP)$ indexed by 
$TMP$.
The index $TMP$ is associated with the main diagonal connecting the origin of the 
$4$-cube (defined by $TMP,T,TX,TY$) with its opposite vertex.

\begin{verbatim}
 for (T=0;T<2;T+=1)
   for (TX=0;TX<2;TX+=1)  
     for (TY=0;TY<2;TY+=1)
        accum = a(T,TX,TY)
        
Introduction of TMP in the clock view:
for (TMP = 0;TMP<16;TMP+=8)
  for (T=TMP;T<TMP+8;T+=4)
    for (TX=T;TX<T+4;TX+=2)  
      for (TY=TX;TY<TX+2;TY+=1)
        s(TMP%8) += a((T-TMP)/4,(TX-T)/2,TY-TX)
\end{verbatim}

Unrolling of $TMP$, leads to two loops, that could be computed in parallel.
\begin{verbatim}
  for (T=0;T<4;T+=4)
    for (TX=T;TX<T+4;TX+=2)  
      for (TY=TX;TY<TX+2;TY+=1)
        s(0)+=a(0,(TX-T)/2,TY-TX)
 
  for (T=4;T<8;T+=4)
    for (TX=T;TX<T+4;TX+=2)  
      for (TY=TX;TY<TX+2;TY+=1)
        s(1)+=a((1,(TX-T)/2,TY-TX)
   
   S=s(0)+s(1)
\end{verbatim}

\subsection{Recurrent patterns}
The lattice is considered as a multi-dimensional $k$-cube (it is usually a
$k$-rectangle, the transition from $k$-cubes to $k$-rectangle is 
straightforward). Our enumeration is based on factorization, i.e representing this cube 
as a product of smaller cubes (with side size limited by a given threshold). 
We start with an "unit" cube, and represent it as a product between
the "unit" cube and a "quotient" cube. This "quotient" cube is represented as a 
product between the "unit" cube" and a second level "quotient" cube. 
This continue recursively until the resulting "quotient" cube is smaller 
or equal with the "unit" cube.
 
The "unit" cube may be viewed as a moving window, the "quotient" cubes representing 
the positions of its instances in an hierarchy  Alternatively, the "unit" cube may be view 
as a fixed buffer in which indexes (and the variables values) are brought incrementally.

The latter view is related to coloring, and it models the fast (but small) levels of 
memory (as shared memory, cache,
registers, etc) in which the data should be brought to perform the computation.   
We may use an hierarchy of "unit" cubes, and associating for each level of the memory 
hierarchy an "unit" cube of the appropriate size.  

As discussed previously, we work in a time-index space, where the "unit" cube is 
mapped to an "unit" clock, representing the cube points as a tree. 
The tree representing a lattice is constructed incrementally by adding
a subtree (representing the "unit" clock) to leaves of an existent tree.
\section{Lexicographic order and partial results}
The starting point of our approach is a computation using a lexicographic order 
enumeration. Every computation (defined by legal formulas), may be brought to this 
form, possibly using additional indexes for partial results in order to preserve 
the dependencies. In this case, additional formulas are added to the original 
formulas, exposing the use of the new introduced indexes. 

In a sense, this is very similar with the accumulator unfolding discussed in the previous 
section. The difference is that in case of more complex dependencies, the introduction
of new formulas is motivated by the need to introduce a lexicographic order
complying with data dependencies.
We will exemplify the above ideas for a stencil computation, matrix transpose and 
matrix multiplication. 
  
\subsection{Stencil computation}
Coloring for stencils computation have been long time advocated, 
but performed only for the Jacobi stencil using two colors (the 
black and red scheme).  
Our coloring algorithm works with any stencil 
computation, the minimal number of colors required being determined by the 
stencil structure. As a difference from the cases previously considered, 
in the stencils formulas have displacements different from zero.
 
We consider the stencil formula $a(I,J)+=a(I,J+1)+
a(I+1,J)+a(I+1,J+1)$ to be performed on the lattice (grid) spawned by the
indexes $I \times J$. To simplify only the case with all displacements 
non-negative are considered (but it is possible to extend supporting to 
negative displacements).

In this formulation, we need to be careful when advancing on the lattice 
defined by $I$ and $J$ to not override a location $a(I,J)$ while the
value stored may have a further use (the problem was discussed in a 
previous section and was the rationale behind the lexicographic order
requirement). The solution for avoiding the above problem is to use
different memory spaces for the result and the operands of the 
stencil pattern.  

The length of paths from stencil origin to stencil points (given by the 
sum of displacements) are used to structure the enumeration 
and allocate the temporary space required for the lexicographic order.

For our examples, two time graduations $S$ and $T$ (corresponding to 
different displacements of the stencil points) are considered.  
We consider the $4$-clock $16,8,4,2$ and the mapping
$S \rightarrow 16, I \rightarrow 8, T \rightarrow 4, J \rightarrow 2$.
This clock is further structured as a $2$-clock with graduations $S$ 
and $T$, and $I$ and $J$ reached via convolutions. 

To represent the computation, the $form$ (multiple $for$) construct  
is used. 
\begin{verbatim}
Pattern:
a(I,J)+=a(I,J+1)+a(I+1,J)+a(I+1,J+1)

form [(S=0;S<2;S+=2)
     [(I=0;I<2;I+=1)
     [(T=0;T<2;T+=2)
     [(J=0;J<2;J+=1)
        a(I,J)=a(I+S%2,J+T%2)
               a(I+(S+1)%2,J+T%2)
               a(I+S%2,J+(T+1)%2)
               a(I+(S+1)%2,J+(T+1)%2)
\end{verbatim}

Introducing a first level of convolutions.
\begin{verbatim}
form[[for (S=0;S<2;S+=2)
      for (I=S;I<S+2;I+=1)]
     [for (T=0;T<2;T+=2)
      for (J=T;J+T<2;J+=1)]]
        a(I-S,J-T)=a((I-S)+S%2,(J-T)+T%2)
                   a((I-S)+(S+1)%2,(J-T)+T%2)
                   a((I-S)+S%2,(J-T)+(T+1)%2)
                   a((I-S)+(S+1)%2,(J-T)+(T+1)%2)
\end{verbatim}        
 
Introducing a second level of convolutions.
\begin{verbatim}      
for (S=0;S<32;S+=16)
  for (I=S;I<S+16;I+=8)
    for (T=I+S;T<I+S+8;T+=4)
      for (J=T;J<T+4;J+=2)
        a(I-S,J-T-S)=a((I-S)/8+S%16,(J-T-S)/2+T%4)
                     a((I-S)/8+(S+16)%16,(J-T-S)/2+T%4)
                     a((I-S)/8+S%16,(J-T-S)/2+(T+4)%4)
                     a((I-S)/8+(S+4)%16,(J-T-S)/2+(T+4)%4)
\end{verbatim}                  
                   
The memory allocated for stencil computation, may be structured according to coloring
(assigning contiguous locations for points of same color) as a separate step before beginning the
computation. For a sufficient great number of iterations, the benefits due to better 
locality for each stencil iteration, will overhead the cost of restructuring.

\subsection{Matrix transpose}
The computation considered have cycles (of size greater then one) 
between result and operand indexes. 
For instance, transpose $a(I,J) = a(J,I)$ induce a cycle of size two,
$I \rightarrow J \rightarrow I$. This cycle is constructed 
by considering indexes in the same position in result and 
operands. Until now we considered only cycles of size one,
$I \rightarrow I$. As we will show in the example below, the cycle 
of size two is "unraveled" by allocating to it a new index. 

We consider the initial computation, for a cube of dimension 2. 
\begin{verbatim}
Computational pattern:
   a(I,J)=a(J,I)

   form [(I=0;I<2;I+=1)
         (J=0;J<I;J+=1)
           tmp=a(I,J)
           a(I,J)=a(J,I) 
           a(I,J)=tmp
\end{verbatim}

An additional index $T$ is added, and the unfolded $3$-cube
is shown below. Two memory locations $tmp(2)$ are used for 
the temporary results.
\begin{verbatim}
   form [(T=0;T<2;T+=1)
         (I=0;I<2;I+=1)
         (J=0;J<I;J+=1)]
       tmp(T)=a(I,J)
       a(I,J)=a(J,I)
       a(J,I)=tmp(T) 
\end{verbatim}

We consider a clock $(8,4,2)$.  
Mapping is $T \rightarrow 8$, $I \rightarrow 4$,
$J \rightarrow 2$. 
The clock based representation is shown below.  

\begin{verbatim}
  for (T=0;T<8;T+=4)
    for (I=T;I<T+4;I+=2)
      // Transformation is J->J+2, instead of J->I+J (already J<I).
      for (J=2;J<I+2;J+=1)
        tmp(T%4)=a((I-T)/2,J-2)
        a((I-T)/2,J-2) = a(J-2,(I-T)/2)
        a(J-2,(I-T)/2)=tmp(T%4)
\end{verbatim}

The resulting loop representation may be view as a parallelization of 
$T = a(I,J),a(I,J) = a(J,I),a(J,I) = T$, standard way of representing
interchange. We note that the amount of parallelism is function
of temporary locations number.

We denote the body of the loop by $B(T,I,J)$. 
Unrolling the temporary dimension results in two copies of the body $B(0,I,J)$ 
and $B(1,I,J)$ could be performed in parallel as disjunct sets of I and J are used.
Each copy uses one of the two temporary locations $tmp(0)$ and $tmp(1)$.

\begin{verbatim} 
  B(0,I,J)
  for (I=0;I<4;I+=2)
    for (J=2;J<I+2;J+=1)
      tmp(0)= a((I/2,J-2)
      a((I-T)/2,J-2) = a(J-2,(I-T)/2)
      a(J-2,(I-T)/2) = tmp(T/4)
   
   B(1,I,J) 
   for (I=4;I<8;I+=2)
     for (J=2;J<I+2;J+=1)
       tmp(1)= a((I-4)/2,J-2)
       a((I-4)/2,J-2) = a(J-2,(I-4)/2)
       a(J-2,(I-4)/2) = tmp(1)  
\end{verbatim}

\subsection{Join product}
We consider two sets of indexes $I_{m}$, respective $J_{n}$ and the index space 
defined by their union.
Until now we assumed that these sets of indexes are disjunct. In this subsection, 
a non-empty set of indexes $K_{r}$ may be shared between sets of indexes
(i.e. it is their intersection). 
The two sets of indexes may be rewritten as $I_{m} = I_{p} \cup K_{r}$
and $J_{n} = J_{q} \cup K_{r}$, the sets $I_{p}$ and $J_{q}$ being disjunct.

We want to compute the $join$ product defined on $I_{p} \times I_{q}$ eliminating 
the shared indexes $K_{r}$ from the index space $I_{m} \times I_{n}$, and having the
property that $join(I_{p},I_{q}) = (I_{p},K_{r})(K_{r},J_{p})$ for every index in
$I_{p},I_{q},K_{r}$. 

Matrix multiplication is a simple case, where each of the sets of indexes
$I_{p},I_{q},I_{r}$ consists of a single index $I,J,K$. The clock view offers an 
efficient way for computing (enumerating) the join product of indexes $I$ and $J$, 
via a diagonal embedding defined by the shared index $K$. We present this for the 
case of matrix multiplication, but the approach may be generalized to sets 
$I_{p},I_{q},I_{r}$ with arbitrary number of indexes. 

Every matrix may be view as a $2$-cube (the result may be extended to   
but the transition from cube to rectangle is straightforward), and the matrix
multiplication becomes the product of two $2$-cubes having a shared index. 
The same clock is associated to each cube, with the shared index mapped to
time index index, and the remaining indexes mapped to clock convolutions. 

The usual code for matrix multiplications of $2$-cubes of size $2$
using the $for$ construct.  
\begin{verbatim} 
  for (I=0;I<2;I++)
    for (J=0;J<2;I++)
      for (K=0;K<2;K++)
        a(I,J)+=b(I,K)*c(K,J)
\end{verbatim}

Using the equivalent $form$ (multiple $for$) construct.
\begin{verbatim} 
  form [(K=0;K<2;K+=1)
        (I=0;I<2;I+=1)
        (J=0;J<2;J+=1)]
           a(I,J)+=b(I,K)*c(K,J)
\end{verbatim}

Mapped to the $2$-clock $(8,4)$ with $K \rightarrow 8,(I,J) \rightarrow 4$.
\begin{verbatim} 
  for (K=0;K<8;K+=4)
     form [(I=K;I<K+2;I+=1)
           (J=K;J<K+2;J+=1)]
              a(I-K,J-K)+=b(I-K,K)*c(K,J-K)
\end{verbatim}

Mapped to the $3$-clock $(8,4,2)$ with $K \rightarrow 8,I \rightarrow 4, J \rightarrow 2$. 
\begin{verbatim} 
  for (K=0;K<8;K+=4)
    for (I=K;I<K+4;I+=2)
      for (J=I;J<K+2;J+=1)]
        a(I-K,J-I)+=b(I-K,K)*c(K,J-I)
\end{verbatim}

We denote the formula in the body of the loop by $B(K,I,J)$. The unfolding of $K$ defines 
a tree of formulas. 
After unfolding of $K$, $B(0,I,J)$ and $B(1,I,J)$ may be computed in parallel.

\section{Sparse lattices and graphs}
The algorithms previously presented may be used for graph based computations,
as they describe enumerations of a graph vertexes. The graph is represented 
by incidence relations that may be implemented using a dense or sparse 
representation.

For sparse representations, the lexicographic order (and iterative view) 
is replaced by the depth-first order. Like for the dense case, the enumeration 
is structured by paths with length sums of powers of two. 

In the dense case, the mapping of the graph vertexes to a clock is performed  
statically, but for the sparse case this mapping is performed dynamically as 
the enumeration proceed. In this way, irregular graph structure (and the 
mapping to them) are "fitted" to a clock structure. The dynamic mapping may 
affect the shape of recursive patterns, the sparse structure being less 
regular then the dense one.  

Similar transformations are done for computation parallelization, and are 
behind parallel programming patterns (\citet{Ma}), as well as of some parallel 
versions of sequential algorithms. 
      
\bibliographystyle{plainnat}
\bibliography{biblio}

\end{document}